# CHAPTER 9

# Multiscale Discrete Dislocation Dynamics Plasticity

H.M. Zbib, M. Hiratani[*] and M. Shehadeh

School of Mechanical and Materials Engineering
Washington State University, Pullman, WA 99164-2920.
(509) 335-7832, zbib@wsu.edu

## 9.1 Introduction

Deformation and strength of crystalline materials are determined to a large extent by underlying mechanisms involving various crystal defects, such as vacancies, interstitials and impurity atoms (point defects), dislocations (line defects), grain boundaries, heterogeneous interfaces and microcracks (planar defects), chemically heterogeneous precipitates, twins and other strain-inducing phase transformations (volume defects). Most often, dislocations define plastic yield and flow behavior, either as the dominant plasticity carriers or through their interactions with the other strain-producing defects.

Dislocation as a line defect in a continuum space was first introduced as a mathematical concept in the early 20th century by Voltera (1907) and Somigliana (1914). They considered the elastic properties of a cut in a continuum, corresponding to slip, disclinations, and/or dislocations. But associating these geometric cuts to dislocations in crystalline materials was not made until the year 1934. In order to explain the less than ideal strength of crystalline materials, Orowan (1934), Polanyi (1934) and Taylor (1934) simultaneously hypothesized the existence of dislocation as a crystal defect. Later in the late 50's, the existence of dislocations was experimentally confirmed by Hirsch, et al. (1956) and Dash (1957). Presently these crystal defects are routinely observed by various means of electron microscopy.

---

[*] Current address: Lawrence Livermore National Laboratory, L-353  P.O.Box 808
7000 East Av. Livermore, CA 94551

A dislocation can be easily understood by considering that a crystal can deform *irreversibly* by slip, i.e. shifting or sliding along one of its atomic planes. If the slip displacement is equal to a lattice vector, the material across the slip plane will preserve its lattice structure and the change of shape will become permanent. However, rather than simultaneous sliding of two half-crystals, slip displacement proceeds sequentially, starting from one crystal surface and propagating along the slip plane until it reaches the other surface. The boundary between the slipped and still unslipped crystal is a dislocation and its motion is equivalent to slip propagation. In this picture, crystal plasticity by slip is a net result of the motion of a large number of dislocation lines, in response to applied stress. It is interesting to note that this picture of deformation by slip in crystalline materials was first observed in the nineteenth century by Mügge (1883) and Ewing and Rosenhain (1899). They observed that deformation of metals proceeded by the formation of slip bands on the surface of the specimen. Their interpretation of these results was obscure since metals were not viewed as crystalline at that time.

Over the past seven decades, experimental and theoretical developments have firmly established the principal role of dislocation mechanisms in defining material strength. It is now understood that macroscopic properties of crystalline materials are derivable, at least in principle, from the behavior of their constituent defects. However, this fundamental understanding has not been translated into a continuum theory of crystal plasticity based on dislocation mechanisms. The major difficulty in developing such a theory is the multiplicity and complexity of the mechanisms of dislocation motion and interactions that make it impossible to develop a quantitative analytical approach. The problem is further complicated by the need to trace the spatiotemporal evolution of a very large number of interacting dislocations over very long periods of time, as required for the calculation of plastic response in a representative volume element. Such practical intractability of the dislocation-based approaches, on one hand, and the developing needs of material engineering at the nano and micro length scales on the other, have created the current situation when equations of crystal plasticity used for continuum modeling are phenomenological and largely disconnected from the physics of the underlying dislocation behavior.

Bridging the gap between dislocation physics and continuum crystal plasticity has become possible with the advancement in computational technology with bigger and faster computers. To this end, over the past two decades various discrete dislocation dynamics models have been developed. The early discrete dislocation models were two-dimensional (2D) and consisted of periodic cells containing multiple dislocations whose behavior was governed by a set of simplified rules (Lepinoux and Kubin 1987; Ghoniem and Amodeo 1988; Groma and Pawley 1993; Van der Giessen and Needleman 1995; Wang and LeSar 1995; Le and Stumpf 1996). These simulations, although served as a useful conceptual framework, were limited to 2D and, consequently, could not directly account for such important features in dislocation dynamics as slip geometry, line tension effects, multiplication, dislocation intersection and cross-slip, all of which are crucial for the formation of dislocation patterns. In the 90's, development of new computational approaches of dislocation dynamics (DD) in three-dimensional space (3D) generated hope for a principal breakthrough in our current understanding of dislocation mechanisms and their connection to crystal plasticity (Kubin and Canova 1992; Canova, Brechet et al. 1993; Hirth, Rhee et al. 1996; Zbib, Rhee et al. 1996). In these new models, dislocation motion and interactions with other defects, particles and surfaces are explicitly considered. However, complications with respect to dislocation multiplications, self-interactions and interactions with other defects, and keeping track of complex mechanisms and reactions have provided a new set of challenges for developing efficient computational algorithms.

The DD analysis and its computer simulation modeling devised by many researchers (Ghoniem and Amodeo 1988; Canova, Brechet et al. 1992; Kubin 1993; Schwarz and Tersoff 1996; Zbib, Rhee et al. 1996; Zbib, Rhee et al. 1998) has been advanced significantly over the past decade. This progress has been further magnified by the idea to couple DD with continuum mechanics analysis in association with computational algorithms such as finite elements. This coupling may pave the way to better understanding of the local response of materials at the nano and micro scales and globally at the macroscale (Zbib and Diaz de la Rubia 2002), increasing the potential for future applications of this method in material, mechanical, structural and process engineering analyses. In the following, the principles of DD analysis will be presented followed by the procedure for the measurement of local quantities such as plastic distortion and internal stresses. The incorporation of DD technique into the three-dimensional plastic continuum mechanics-based finite elements modeling will then be described. Finally, examples are provided to illustrate the applicability of this powerful technique in material engineering analysis.

## 9.2 Theoretical Fundamentals of the Method

In order to better describe the mathematical and numerical aspects of the DD methodology, first we will identify the basic geometric conditions and kinetics that control the dynamics of dislocations. This will be followed by discussion of the dislocation equation of motion, elastic interaction equations, and descritization of these equations for numerical implementation.

### 9.2.1 Kinematics and Geometric Aspects

A dislocation is a line defect in an otherwise perfect crystal described by its line sense vector $\xi$ and Burgers vector *b*. The Burgers vector has two distinct components: edge, perpendicular to its line sense vector, and screw, parallel to its line sense vector. Under loading, dislocations glide and propagate on slip planes causing deformation and change of shape. When the local line direction becomes parallel to the Burgers vector, the dislocation may propagate into other slip planes. This switching of the slip plane, which makes the motion of dislocations three-dimensional, and is better known as cross slip is an important recovery phenomena to be dealt with in dislocation dynamics. In addition to glide and cross slip, dislocations can also climb in a non-conservative three-dimensional motion by absorbing and/or emitting intrinsic point defects, vacancies, and interstitials. Some of these phenomena become important at high loads level or temperatures when point defects become more mobile. In summary, the 3D dislocation dynamics accounts for the following geometric aspects:

- Dislocation topology; 3D geometry, Burgers vector and line sense.
- Identification of all possible slip planes for each dislocation.
- Changes in the dislocation topology when part of it cross-slips and or climbs to another plane.
- Multiplication and annihilation of dislocation segments.
- Formation of complex connections and intersections such as junctions, jogs and branching of the dislocation in multiple directions.

## 9.2.2 Kinetics and Interaction Forces

The dynamics of the dislocation is governed by a "*Newtonian*" equation of motion, consisting of an inertia term, damping term, and driving force arising from short-range and long-range interactions. Since the strain field of the dislocation varies as the inverse of distance from the dislocation core, dislocations interact among themselves over long distances. As the dislocation moves, it has to overcome internal drag, and local barriers such as the Peirels stresses. The dislocation may encounter local obstacles such as stacking fault tetrahedra, defect clusters and vacancies that interact with the dislocation at short ranges and affect its local dynamics. Furthermore, the internal strain field of randomly distributed local obstacles gives rise to stochastic perturbations to the encountered dislocations, as compared with deterministic forces such as the applied load. This stochastic stress field also contributes to the spatial dislocation patterning in the later deformation stages. Therefore, the strain field of local obstacles adds spatially irregular uncorrelated noise to the equation of motion. In addition to the random strain fields of dislocations or local obstacles, thermal fluctuations also provide a stochastic source in dislocation dynamics. Dislocations also interact with free surfaces, cracks, and interfaces, giving rise to what is termed as image forces. In summary, the dislocation may encounter the following set of forces:

- Drag force, $B\boldsymbol{v}$, where $B$ is the drag coefficient and $\boldsymbol{v}$ is the dislocation velocity.
- Peirels stress $\boldsymbol{F}_{Peirels}$.
- Force due to externally applied loads, $\boldsymbol{F}_{external}$.
- Dislocation-dislocation interaction force $\boldsymbol{F}_D$.
- Dislocation self-force $\boldsymbol{F}_{self}$.
- Dislocation-obstacle interaction force $\boldsymbol{F}_{obstacel}$.
- Image force $\boldsymbol{F}_{image}$.
- Osmotic force $\boldsymbol{F}_{Osmatic}$ resulting from non-conservative motion of dislocation (climb) and results in the production of intrinsic point defects.
- Thermal force $\boldsymbol{F}_{thermal}$ arising from thermal fluctuations.

The DD approach attempts to incorporate all of the aforementioned kinematics and kinetics aspects into a computational traceable framework. In the numerical implementation, three-dimensional curved dislocations are treated as a set of connected segments as illustrated in Figure 1. It is possible to represent smooth dislocations with any desired degree of realism, provided that the discretization resolution is taken high enough for accuracy (limited by the size of the dislocation core radius $r_0$, typically the size of one Burgers vector). In such a representation, the dynamics of dislocation lines is reduced to the dynamics of discrete degrees of freedom of the dislocation nodes connecting the dislocation segments.

### 9.2.3 Dislocation Equation of Motion

The velocity $v$ of a dislocation segment $s$ is governed by a first order differential equation consisting of an inertia term, a drag term and a driving force vector (Hirth 1992; Hirth, Zbib et al. 1998; Huang, Ghoniem et al. 1999), such that

$$m_s \dot{v} + \frac{1}{M_s(T,p)} v = F_s \quad \text{with} \quad m_s = \frac{1}{v}\left(\frac{dW}{dv}\right) \tag{1}_1$$

$$F_s = F_{Peirels} + F_D + F_{Self} + F_{External} + F_{Obstacle} + F_{Image} + F_{Osmotic} + F_{Thermal} \tag{1}_2$$

In the above equation the subscript $s$ stands for the segment, $m_s$ is defined as the effective dislocation segment mass density, $M_s$ is the dislocation mobility which could depend both on the temperature $T$ and the pressure $P$, and $W$ is the total energy per unit length of a moving dislocation (elastic energy plus kinetic energy). As implied by (1)$_2$, the glide force vector $F_s$ per unit length arises from a variety of sources described in the previous section. The following relations for the mass per unit dislocation length have been suggested (Hirth, Zbib et al. 1998) for screw $(m_s)_{screw}$ and edge $(m_s)_{edge}$ dislocations when moving at a high speed.

$$(m_s)_{screw} = \frac{W_0}{v^2}(-\gamma^{-1} + \gamma^{-3})$$

$$(m_s)_{edge} = \frac{W_0 C^2}{v^4}(-16\gamma_l - 40\gamma_l^{-1} + 8\gamma_l^{-3} + 14\gamma + 50\gamma^{-1} - 22\gamma^{-3} + 6\gamma^{-5})$$

(2)

where $\gamma_l = (1 - v^2/C_l^2)^{\frac{1}{2}}$, $\gamma = (1 - v^2/C^2)^{\frac{1}{2}}$, $C_l$ is the longitudinal sound velocity, $C$ is the transverse sound velocity, $\nu$ is Poisson's ratio, $W_0 = \frac{Gb^2}{4\pi} ln(R/r_0)$ is the rest energy for the screw per unit length, $G$ is the shear modulus. The value of $R$ is typically equal to the size of the dislocation cell (about 1000 $b$), or in the case of one dislocation is the shortest distance from the dislocation to the free surface (Hirth and Lothe 1982). In the non-relativistic regime when the dislocation velocity is small compared to the speed of sound, the above equations reduce to the familiar expression $m = \beta \rho b^2 ln(R/r_0)$, where $\beta$ is a constant dependent on the type of the dislocation, and $\rho$ is the mass density.

*9.2.3.a Dislocation Mobility Function*

The reliability of the numerical simulation depends critically on the accuracy of the dislocation drag coefficient $B$ (= $1/M$) that is material dependent. There are a number of phenomenological relations for the dislocation glide velocity $v_g$ (Kocks, Argon et al. 1975; Sandstrom 1977), including relations of power law forms and forms with an activation term in an exponential or as the argument of a sinh form. Often, however (Johnston and Gilman 1959; Sandstrom 1977) the simple power law form is adopted for expedience, e. g. $v_g = v_s(\tau_e/\tau_s)^m$, resulting in nonlinear dependence of $M$ on the stress. In a number of cases of pure phonon/electron damping control or of glide over the Peierls barrier a constant mobility (with $m=1$), predicts the results very well. This linear form has been theoretically predicted for a number of cases as discussed by Hirth and Lothe (1982).

Mechanisms to explain dislocation drag have been studied for long time and the drag coefficients have been estimated in numerous experimental and theoretical works by atomistic simulations or quantum mechanical calculations [see, for example, the review by Al'shitz (1992)]. The determination of each of the two components (phonon and electron drag) that constitute the drag coefficient for a specific material is not trivial, and various simplifications have been made, e.g. the Debye model neglects Van Hove singularities in phonon spectrum (Ashcroft and Mermin 1976), isotropic approximation of deformation potentials, and so on. Also the values are sensitive to various parameters such as the mean free path or core radius. Nevertheless, in typical metals, the phonon drag $B_{ph}$ range is 30 ~ 80 $\mu$Pa.s at room temperature and less than 0.1 $\mu$Pa.s at very low temperatures around 10K, while for the electron drag $B_e$ the range is a few $\mu$Pa.s and expected to be temperature independent. Under strong magnetic fields at low temperature, macroscopic dislocation behavior can be highly sensitive to orientation relative to the field within accuracy of 1% (McKrell and Galligan 2000). Except for special cases such as deformation under high strain rate, weak dependences of drag on dislocation velocity are usually neglected.

Examples of temperature dependence of each component of the drag coefficient can be found for the case of edge dislocation in copper (Hiratani and Nadgorny 2001), or in Molybdenum (Jinpeng, Bulatov et al. 1999). Generally, however, the dislocation mobility could be, among other things, a fuction of the angle between the Burgers vector and the dislocation line sense, especially at low temperatures. For example, Wasserbäch (1986) observed that at low deformation temperatures (77 to 195 K) the dislocation structure in Ta single crystals consisted of primary and secondary screw dislocations and of tangles of dislocations of mixed characters, while at high temperatures (295 to 470 K) the behavior was similar to that of fcc crystals. In the work of Mason and MacDonald (1971) they measured the mobility of dislocation of an unidentified type in NB as $4.2 \times 10^4$ (Pa.s)$^{-1}$ near room temperature. A smaller value of $3.3 \times 10^3$ (Pa.s)$^{-1}$ was obtained by Urabe and Weertman (1975) for the mobility of edge dislocation in Fe. The mobility for screw dislocations in Fe was found to be about a factor of two smaller than that of edge dislocations near room temperature. A theoretical model to explain this large difference in behavior is given in Hirth and Lothe (1982) and is based on the observation that in bcc materials the screw dislocation has a rather complex three-dimensional core structure, resulting in a high Peirels stress, leading to a relatively low mobility for screw dislocations while the mobility of mixed dislocations is very high.

*9.2.3.b Dislocation Collisions*

When two dislocations collide, their response is dominated by their mutual interactions and becomes much less sensitive to the long-range elastic stress associated with external loads, boundary conditions, and all other dislocations present in the system. Depending on the shapes of the colliding dislocations, their approach trajectories and their Burgers vectors, two dislocations may form a dipole, or react to annihilate, or to combine to form a junction, or to intersect and form a jog. In the DD analysis, the dynamics of two colliding dislocations is determined by the mutual interaction force acting between them. In the case that the two dislocation segments are parallel (on the same plane and or intersecting planes) and have the same Burgers vector with opposite sign they would annihilate if the distance between them is equal to the core size. Otherwise, the colliding dislocations would align themselves to form a dipole, a jog or a junction depending on their relative position. A comprehensive review of short-range interaction rules can be found in Rhee, Zbib et al. (1998).

*9.2.3.c Discretization of Dislocation Equation of Motion*

Equation (1) applies to every infinitesimal length along the dislocation line. In order to solve this equation for any arbitrary shape, the dislocation curve may be discretized into a set of dislocation segments as illustrated in Figure 1. Then the velocity vector field over each segment may be assumed to be linear and, therefore, the problem is reduced to finding the velocity of the nodes connecting these segments. There are many numerical techniques to solve such a problem. Consider, for example, a straight dislocation segment *s* bounded by two nodes *j* and *j+1* as depicted in Figure 1. Then within the finite element formulation (Bathe 1982), the velocity vector field is assumed to be linear over the dislocation segment length. This linear vector field *v* can be expressed in terms of the velocities of the nodes such that $v = [N^D]^T V^D$ where $V^D$ is the nodal velocity vector and $[N^D]$ is the linear shape function vector (Bathe 1982). Upon using the Galerkin method, equation (1) for each segment can be reduced to a set of six equations for the two discrete nodes (each node has three degrees of freedom). The result can be written in the following matrix-vector form.

$$[M^D]\dot{V}^D + [C^D]V^D = F^D \tag{3}$$

where

$[M^D] = m_s \int [N^D][N^D]^T dl$     is the dislocation segment 6x6 mass matrix,

$[C^D] = (1/M_s) \int [N^D][N^D]^T dl$     is the dislocation segment 6x6-damping matrix, and

$F^D = \int [N^D] F_s dl$     is the 6x1 nodal force vector.

Then, following the standard element assemblage procedure, one obtains a set of discrete system of equations, which can be cast in terms of a global dislocation mass matrix, global dislocation damping matrix, and global dislocation force vector. In the case of one dislocation loop and with ordered numbering of the nodes around the loop, it can be easily shown that the global matrices are banded with half-bandwidth equal to one. However, when the system contains many loops that interact among themselves and new nodes are generated and/or annihilated continuously, the numbering of the nodes becomes random and the matrices become un-banded. To simplify the computational effort, one can employ the lumped mass and damping matrix method. In this method, the mass matrix $[M^D]$ and damping matrix $[C^D]$ become diagonal matrices (half-bandwidth equal to zero), and therefore the only coupling between the equations is through the nodal force vector $F^D$. The computation of each component of the force vector is described below.

### 9.2.4 The Dislocation Stress and Force Fields

The stress induced by any arbitrary dislocation loop at an arbitrary field point $p$ can be computed by the Peach-Koehler integral equation given in Hirth and Lothe (1982). This integral equation, in turn, can be evaluated numerically over many loops of dislocations by discretizing each loop into a series of line segments. If we denote,

$N_l$ = total number of dislocation loops
$n_s^{(l)}$ = number of segments of loop $l$

$n_n^{(l)}$ = number of nodes associated with the segments of loop $n_n^{(l)} = n_s^{(l)} + 1$

$N_s$ = total number of segments = $n_s^{(l)} \times N_l$

$N_n$ = total number of nodes = $n_n^{(l)} \times N_l$

$l_s$ = length of segment $s$

$r$ = distance from point $p$ to the segment $s$

Then the discretized form of the Peach-Koehler integral equation for the stress at any arbitrary field point $p$ becomes

$$\overset{d}{\sigma}_{ij}(p) = \sum_{l=1}^{N_l} \sum_{s=1}^{N_s^{(l)}} \left\{ -\frac{G}{8\pi} \int_{l_s} b_p \in_{mpi} \frac{\partial}{\partial x'_m} \nabla'^2 R \, dx'_j - \frac{G}{8\pi} \int_{l_s} b_p \in_{mpi} \frac{\partial}{\partial x'_m} \nabla'^2 R \, dx'_j \right. \\ \left. - \frac{G}{4\pi(-v)} \int_{l_s} b_p \in_{mpk} \left( \frac{\partial^3 R}{\partial x'_m \partial x'_i \partial x'_j} - \delta_{ij} \frac{\partial}{\partial x'_m} \nabla'^2 R \right) dx'_k \right\} \quad (4)$$

where $\in_{ijk}$ is the permutation symbol. The integral over each segment can be explicitly carried out using the linear element approximation. Exact solution of equation (4) for a straight dislocation segment can be found in DeWit (1960) and Hirth and Lothe (1982). However, evaluation of the above integral requires careful consideration as the integrand becomes singular in cases where point $p$ coincides with one of the nodes of the segment that integration is taken over, i.e., self-segment integration. Thus,

- If $p$ is not part of the segment $s$, there is no singularity since $r \neq 0$ and the ordinary integration procedure may be performed.
- If $p$ coincides with a node of the segment $s$ where the integration should be carried out, special treatment is required due to the singular nature of the stress field as $r \to 0$. Here, the regularization scheme developed by Zbib and co-workers have been employed.

In general, the dislocation stresses can be decomposed into the following form.

$$\overset{d}{\sigma}(p) = \sum_{s=1}^{N_s - 2} \overset{d}{\sigma}^{(s)} + \overset{d}{\sigma}^{(p+)} + \overset{d}{\sigma}^{(p-)} \quad (5)$$

where $\overset{d}{\boldsymbol{\sigma}}{}^{(s)}$ is the contribution to the stress at point $p$ from a segment $s$, and $\overset{d}{\boldsymbol{\sigma}}{}^{(p+)}$, $\overset{d}{\boldsymbol{\sigma}}{}^{(p-)}$ are the contributions to the stress from the two segments that are shared by a node coinciding with $p$ which will be further discussed below.

Once the dislocation stress field is computed the forces on each dislocation segment can be calculated by summing the stresses along the length of the segment. The stresses are categorized into those coming from the dislocations as formulated above and also from any other external applied stresses plus the internal friction (if any) and the stresses induced by any other defects. A model for the osmotic force $\boldsymbol{F}_{\text{Osmotic}}$ is given in Raabe (1998) and its inclusion in the total force is straightforward since it is a deterministic force. However, the treatment of the thermal force $\boldsymbol{F}_{\text{thermal}}$ is not trivial since this force is stochastic in nature, requiring a special consideration and algorithm leading to what is called *Stochastic Dislocation Dynamics* (SDD) as developed by Hiratani and Zbib (2002). Therefore, the force acting on each segment can be written as:

$$\boldsymbol{F}_s = \sum_{m=1}^{N_s} (\overset{d}{\boldsymbol{\sigma}}{}^{(m)} + \overset{a}{\boldsymbol{\sigma}}{}^{(m)} + \boldsymbol{\tau}_s ).\boldsymbol{b}_s \times \boldsymbol{\xi}_s = \overset{d}{\boldsymbol{F}}_s + \overset{a}{\boldsymbol{F}}_s + \boldsymbol{F}_{\text{thermal}} \tag{6}$$

where $\overset{d}{\boldsymbol{\sigma}}{}^{(m)}$, is the contribution to the stresses along segment $m$ from all the dislocations (dislocation-dislocation interaction), $\overset{a}{\boldsymbol{\sigma}}{}^{(m)}$ is the sum of all externally applied stresses, internal friction (if any) and the stresses induced by any other defects, and $\boldsymbol{\tau}_s$ is the thermal stress; $\overset{d}{\boldsymbol{F}}_s$, $\overset{a}{\boldsymbol{F}}_s$ and $\boldsymbol{F}_{\text{thermal}}$ are the corresponding total Peach-Koehler (PK) forces.

Using equations (5), the force $\overset{d}{\boldsymbol{F}}_s$ can also be decomposed into two parts one arising from all dislocation segments and one from the self-segment, which is better known as the self-force, that is,

$$\overset{d}{\boldsymbol{F}}_s = \sum_{m=1}^{N_s-1} \overset{d}{\boldsymbol{F}}_s{}^{(m)} + \overset{d}{\boldsymbol{F}}_s{}^{(self)} \tag{7}$$

where $\overset{d}{F}_s{}^{(m)}$ and $\overset{d}{F}_s{}^{(self)}$ are respectively, the contribution to the force on segment *s* from segment *m* and the self-force. In order to evaluate the self-force, a special numerical treatment as given by Zbib, Rhee et al. (1996) and Zbib and Diaz de la Rubia (2002) should be used in which exact expressions for the self-force are given. This approximation works well in terms of accuracy and numerical convergence for segment lengths as small as 20 b. For finer segments, however, one can use a more accurate approximation as suggested by Scattergood and Bacon (1975). Another treatment has been given by Gavazza and Barnett (1976) and used in the recent work of Ghoniem and Sun (1999).

The direct computation of the dislocation forces discussed above requires the use of a very fine mesh, especially when dealing with problems involving dislocation-defect interaction. As a rule to capture the effect of the very small defects, the dislocation segment size must be comparable to the size of the defect. Alternatively, one can use large dislocation segments compared to the smallest defect size, provided that the force interaction is computed over a many points (Gauss points) over the segment length. In this case, the self-force of segment *s* would be evaluated first. Then the force contribution from other dislocations and defects is calculated by computing the stresses at several Gauss points along the length of the segments. The summation as in (6) would then follow according to:

$$F_s = F_s^{self} + \sum_{m=1}^{N_s-1} \frac{1}{n_g} \sum_{g=1}^{n_g} (\overset{d}{\sigma}{}^{(m)}(p_g) + \overset{d}{\sigma}{}^{(m)}(p_g) + .....) \cdot b_s \times \xi_s \tag{8}$$

where $p_g$ is the Gauss point *g* and $n_g$ is the number of Gauss points along segment *s*. The number of Gauss points depends on the length of the segment. As a rule the shortest distance between two Gauss points should be larger or equal to $2r_o$, i.e. twice the core size.

### 9.2.5 The Stochastic Force and Cross-slip

Thermal fluctuations arise from dissipation mechanism due to collision of dislocations with surrounding particles, such as phonons or electrons. Rapid collisions and momentum transfers result in random forces on dislocations. These stochastic collisions, in turn, can be regarded as time-independent noises of thermal forces acting on the dislocations. Suppose the exertion of thermal forces follows a Gaussian distribution. Then, thermal fluctuations most likely result in very small net forces due to mutual cancellations. However, they sometimes become large and may cause diffusive dislocation motion or thermal activation events such as overcoming obstacle barriers. Therefore, the DD simulation model should also account not only for deterministic effects but also for stochastic forces; leading to a model called "*stochastic discrete dislocation dynamics*" (SDD) (Hiratani and Zbib 2002). The procedure is to include the stochastic force $F_{thermal}$ in the DD model by computing the magnitude of the stress pulse ($\tau_s$) using a Monte Carlo type analysis.

Based on the assumption of the Gaussian process, the thermal stress pulse has zero mean and no correlation (Ronnpagel, Streit et al. 1993; Raabe 1998) between any two different times. This leads to the average peak height given as (Koppenaal and Kuhlmann-Wilsdorf 1964; Hiratani and Zbib 2002)

$$\sigma_s = \sqrt{2kT/M_s \left(b^2 \Delta l \Delta t\right)} \qquad (9)$$

where $k$ denotes Boltzman constant, $T$ absolute temperature of the system, $b$ the magnitude of Burgers vector, $\Delta t$ time step, and $\Delta l$ is the dislocation segment length, respectively. Some values of the peak height are shown in Table 1 for typical combinations of parameters. Here, $\Delta t$ is chosen to be 50 fs, roughly the inverse of the Debye frequency. Although $\Delta l$ or $\Delta t$ does not have a fixed value, such a restriction is imposed so that the system at the energetic global minima should reach thermal equilibrium. The validity of these parameters are checked by measuring the assigned system temperature ($T$) with the kinetic temperature of dislocation of which both ends are fixed, which should coincide on average according to the equipartition law. Although there is no unique way for choosing $\Delta t$ or $\Delta l$ in Eq. (9), the justification can be checked if the segment velocity distribution is correct Maxwellian and if the average kinetic energy of the segment is equivalent to $kT/2$ for each degree of freedom at thermal equilibrium (Hiratani and Zbib 2002). The size of $\Delta l$ may also depend on the microstructure being considered, for example when the size of the local obstacle is in the order of nm, $\Delta l$ is also chosen to be in the same order.

**Table 1**. The stress pulse peak height for various combinations of parameters, $\Delta t = 50$ fs.

| $T$(K) | $1/M(\mu Pa \cdot s)$ | $\tau_h$(MPa) ($\Delta l$ = 5b) | $\tau_h$(MPa) ($\Delta l$ = 10b) |
|---|---|---|---|
| 0 | 2 | 11.5 | 8.11 |
| 50 | 5 | 40.6 | 28.7 |
| 100 | 10 | 81.1 | 57.4 |
| 300 | 30 | 256 | 181 |

Numerical implementation includes an algorithm where stochastic components are evaluated at each time step of which strengths are correlated and sampled from a bivariate Gaussian distribution (Allen and Tildesley 1987)[+]. In the stationary case, distribution of the dislocation segment velocity becomes also a Gaussian with variance of $kT/m^* \Delta l$. Under a constant stress without obstacles, Maxwell distributions around a mean velocity, which is equal to final viscous velocity $b\sigma M$, will be realized as a steady state.

With the inclusion of stochastic forces in DD analysis, one can treat cross-slip (a thermally activated process) in a direct manner, since the duration of waiting time and thermal agitations are naturally included in the stochastic process. For example, for the cross-slip in *fcc* model one can develop a model based on the Escaig-Friedel (EF) mechanism where cross-slip of a screw dislocation segment may be initiated by an immediate dissociation and expansion of Shockley partials. This EF mechanism has been observed to have lower activation energy than Shoeck-Seeger mechanism where the double super kinks are formed on the cross slip plane (this model is

---

[+] Here we generate stress pulses as $\tau_s = \sigma_s \sqrt{-2 \ln r_1} \cos(2\pi r_2)$ where $r_1$ and $r_2$ are uniform random numbers between zero and unity (Allen, M. P. and Tildesley, D. J., 1987).

used for cross-slip in bcc (Rhee, Zbib et al. 1998)). In the EF mechanism, the activation enthalpy $\Delta G$ depends on the interval of the Shockley partials ($d$) and the resolved shear stress on the initial glide plane ($\sigma$). [See, for example, the MD simulation of Rasmussen and Jacobs (1997) and Rao, Parthasarathy et al. (1999)]. The constriction interval $L$ is also dependent on $\sigma$. For example, for the case of copper, the activation energy for cross-slip can be computed using an empirical formula fitted to the MD results of Rao, Parthasarathy et al. (1999).

Figure 2 depicts the $\Delta G(\sigma)$ for the case of copper where the value of the activation free energy is 1.2 eV, and for stacking fault energy is equal to 0.045 J/m$^2$. This activation energy for stress assisted cross-slip is entered as an input data into the DD code. Usually, within the DD code, dislocations are represented as perfect dislocations while a pair of parallel Shockley partials are introduced in the case of screw dislocation segments only for stress calculation. Then a Monte Carlo type procedure is used to select either the initial plane or the cross slip plane according to the activation enthalpy (Rhee, Zbib et al. 1998). For simplicity, one can set the regime of the barrier with area of $L$ x $d$ and strength of $\Delta G / Ld$. The virtual Shockley partials move according to the Langevin forces in addition to the systematic forces according to equation (13) until the partials overcome the barrier and the interval decreases to the core distance. The implementation of this model captures the anisotropic response of cross-slip activation process to the loading direction, and consideration of the time duration (waiting time) during the cross-slip event, which have been missing in the former DD simulations.

### 9.2.6 Modifications for Long-Range Interactions: The Super-Dislocation Principle

Inclusion of the interaction among all the dislocation loops present in a large body is computationally expensive since the number of computations per step would be proportional to $N_s^2$ where $N_s$ is the number of dislocation segments. A numerical compromise technique termed the super-dislocation method, which is based on the multipolar expansion method (Wang and LeSar 1995; Hirth, Rhee et al. 1996; Zbib, Rhee et al. 1996), reduces the order of computation to $N_s \log N_s$ with a high accuracy. In this approach, the dislocations far away from the point of interest are grouped together into a set of equivalent monopoles and dipoles. In the numerical implementation of the DD model, one would divide the 3D computational domain into sub-domains, and the dislocations in each sub-domain (if there are any) are grouped together in terms of monopoles, dipoles, etc. (depending on the desired accuracy) and the their far stress field is then computed.

**9.2.7 Evaluation of Plastic Strains**

The motion of each dislocation segment gives rise to plastic distortion, which is related to the macroscopic plastic strain rate tensor $\dot{\varepsilon}^p$, and the plastic spin tensor $W^p$ via the relations

$$\dot{\varepsilon}^p = \sum_{s=1}^{N_s} \frac{l_s v_{gs}}{2V} (n_s \otimes b_s + b_s \otimes n_s) \tag{10}_1$$

$$W^p = \sum_{s=1}^{N_s} \frac{l_s v_{gs}}{2V} (n_s \otimes b_s - b_s \otimes n_s) \tag{10}_2$$

where $n_s$ is a unit normal to the slip plane, $v_{gs}$ is the magnitude of the glide velocity of the segment, $V$ is the volume of the representative volume element and $N_s = N_l \times n_s^{(l)}$ is the total number of segments. The above relations provide the most rigorous connection between the dislocation motion (the fundamental mechanism of plastic deformation in crystalline materials) and the macroscopic plastic strain, with its dependence on strength and applied stress being explicitly embedded in the calculation of the velocity of each dislocation. Nonlocal effects are explicitly included into the calculation through long-range interactions. Another microstructure quantity, the dislocation density tensor $\alpha$, can also be calculated according to

$$\alpha = \sum \frac{l_s}{V} b_s \otimes \xi_s \tag{11}$$

This quantity provides a direct measure for the net Burgers vector that gives rise to strain gradient relief (bending of crystal) (Shizawa and Zbib 1999).

### 9.2.8 The DD Numerical Solution: An Implicit-Explicit Integration Scheme

An implicit algorithm to solve the equation of motion (3) with a backward integration scheme may be used, yielding the recurrence equation

$$v^{t+\delta t}\left(1+\frac{\Delta t}{m_s M_s}\right)^{t+\delta t} = v^t + \frac{\Delta t}{m_s} F_s^{t+\delta t} \qquad (12)$$

This integration scheme is unconditionally stable for any time step size. However, the DD time step is determined by two factors: i) the shortest flight distance for short-range interactions, and ii) the time step used in the dynamic finite element modeling to be described later. This scheme is adopted since the time step in the DD analysis (for high strain rates) is of the same order of magnitude of the time required for a stable explicit finite element (FE) dynamic analysis. Thus, in order to ensure convergence and stable solution, the critical time $t_c$ and the time step for both the DD and the FE ought to be $t_c = l_c/C_l$, and $\Delta t = t_c / 20$, respectively, where $l_c$ is the characteristic length scale which is the shortest dimension in the finite element mesh.

In summary, the system of equations given in Section 9.2 summarizes the basic ingredients that a dislocation dynamics simulation model should include. There are a number of variations in the manner in which the dislocation curves may be discretized, for example zero order element (pure screw and pure edge), first order element (or piecewise linear segment with mixed character), or higher order nonlinear elements but this is purely a numerical issue. Nonetheless, the DD model should have the minimum number of parameters and, hopefully, all of them should be basic physical and material parameters and not phenomenological ones for the DD result to be predictive. The DD model described above has the following set of physical and material parameters:

- Burgers vectors,
- elastic properties,
- core size (equal to one Burgers vector),
- thermal conductivity and specific heat,

- mass density,
- stacking fault energy, and
- dislocation mobility.

Also there are two numerical parameters: the segment length (minimum segment length can't be less that three times the core size) and the time step (as discussed in Section 9.2.8), but both are fixed to ensure convergence of the result. In the above list, it is emphasized that in general the dislocation mobility is an intrinsic material property that reflects the local drag mechanisms as discussed above. One can use an "effective" mobility that accounts for additional drag from dislocation-point defect interaction, and thermal activation processes if the defects/obstacles are not explicitly impeded in the DD simulations. However, there is no reason not to include these effects explicitly in the DD simulations (as done in the model described above), i.e. dislocation defect interaction, stochastic processes and inertia effects, which actually permits the prediction of the "effective" mobility form the DD analysis (Hiratani and Zbib 2002; Hiratani, Zbib et al. 2002).

## 9.3 Integration of DD and Continuum Plasticity

### 9. 3.1 Continuum Elasto-viscoplasticity

The discrete dislocation model can be coupled with continuum elasto-visoplasticity models, making it possible to correct for *dislocation image stress* and to address a wide range of complex boundary value problems at the microscopic level. In the following, a brief description of this coupling is provided. The coupling is based on a framework in which the material obeys the basic laws of continuum mechanics, i.e. the linear momentum balance.

$$div\,\boldsymbol{\sigma} = \rho\,\dot{\boldsymbol{v}} \tag{13}_1$$

and the energy equation

$$\rho c_v \dot{T} = k \nabla^2 T + \boldsymbol{\sigma} \cdot \dot{\boldsymbol{\varepsilon}}^p \tag{13$_2$}$$

where $\boldsymbol{v} = \dot{\boldsymbol{u}}$ is the particle velocity, $\boldsymbol{u}$, $\rho$, $c_v$ and $k$ are the displacement vector field, mass density, specific heat and thermal conductivity respectively. In DD the representative volume cell analyzed can be further discretized into sub-cells or finite elements, each representing a representative volume element (*RVE*). Then the internal stresses field induced by the dislocations (and other defects) $\boldsymbol{\sigma}^D$ and the plastic strain field within each *RVE* can be calculated at any point within each element. However, and to be consistent with the definition of a *RVE*, the heterogeneous internal stress field can be homogenized over each *RV*E, resulting into an equivalent internal stress $\mathbf{S^D}$ which is homogenous within the *RVE*, i.e.

$$\mathbf{S^D} = <\boldsymbol{\sigma}^D> = \frac{1}{V_{element}} \int_{element} \boldsymbol{\sigma}^D(x) \mathbf{dv} \tag{14}$$

Furthermore, the plastic strain increment results from only the mobile dislocations that exit the *RVE* (or sub-cell) and is computed as in equation (10) but with $V$ being the volume of the element (or sub-cell). The dislocations that are immobile (zero velocity) do not contribute to the plastic strain increment. Therefore, the dislocations within an element induce an internal stress due to their elastic distortion. When some of these dislocations (or all) move and exit the element they leave behind plastic distortion in the element, and the internal stress field should be recomputed by summing the stress from the remaining dislocations in the element. With this homogenization procedure Hooke's law for the *RVE* becomes

$$\boldsymbol{\sigma} + \mathbf{S}^D = [C^e][\varepsilon - \varepsilon^p] \tag{15}$$

where $C^e$ is the fourth-order elastic tensor. In the continuum plasticity theory one would need to develop a phenomenological constitutive law for plastic stress-strain. Here, this ambiguity is resolved by using the explicit expressions given by equation (10)$_1$ for the plastic strain tensor as computed in the dislocation dynamics.

### 9.3.2 Modifications for Finite Domains

The solution for the stress field of a dislocation segment (Hirth and Lothe 1982) is true for a dislocation in an infinite domain and for homogeneous materials. In order to account for finite domain boundary conditions, Van der Giessen and Needleman (1995) developed a 2D model based on the principle of superposition. The method has been extended by Yasin, Zbib et al. (2001) and Zbib and Diaz de la Rubia (2002) to three-dimensional problems involving free surfaces and interfaces as summarized below.

*9.3.2.a Interactions with External Free Surfaces*

In the superposition principle, the two solutions from the infinite domain and finite domain are superimposed. Assuming that the dislocation loops and any other internal defects with self induced stress are situated in the finite domain $V$ bounded by the surface $S$ and subjected to arbitrary external tractions and constraints. Then the stress, displacement, and strain fields are given by the superposition of the solutions for the infinite domain and the actual domain subjected to

$$\sigma = \sigma^\infty + \sigma^*, \quad u = u^\infty + u^*, \quad \varepsilon = \varepsilon^\infty + \varepsilon^* \tag{16}$$

where $\sigma^\infty$, $\varepsilon^\infty$ and $u^\infty$ are the fields caused by the internal defects as if they were in an infinite domain, whereas $\sigma^*$, $\varepsilon^*$ and $u^*$ are the field solutions corresponding to the auxiliary problem satisfying the following boundary conditions

$$\begin{aligned} t &= t^a - t^\infty \quad \text{on } S \\ u &= u^a \quad \text{on part of the boundary } S \end{aligned} \tag{17}$$

where $t^a$ is the externally applied traction, and $t^\infty$ is the traction induced on $S$ by the defects (dislocations) in the infinite domain problem. The traction $-t^\infty = \sigma.n$ on the surface boundary $S$ results into an *image stress* field which is superimposed onto the dislocations segments and, thus, accounting for surface-dislocation interaction.

The treatment discussed above considers interaction between dislocations and external free surfaces, as well as internal free surfaces such as voids. Internal surfaces such as micro-cracks and rigid surfaces around fibers are treated within the dislocation theory framework, whereby each surface is modeled as a pile- up of infinitesimal dislocation loops (Demir, et al. 1993; Demir and Zbib 2001). Hence, defects of these types may be represented as dislocation segments and loops, and their interaction with external free surfaces follows the method discussed above. This subject has been addressed by Khraishi, et al. (2001).

*9.3.2.b Interactions with Interfaces*

The framework described above for dislocations in homogenous materials can be implemented into a finite element code. The model can also be extended to the case of dislocations in heterogeneous materials using the concept of superposition as outlined by Zbib and Diaz de la Rubia (2002). For bi-materials, suppose that domain $V$ is divided into two sub-domains $V_1$ and $V_2$ with domain $V_1$ containing a set of dislocations. The stress field induced by the dislocations and any externally applied stresses in both domains can be constructed in terms of two solutions such that

$$\boldsymbol{\sigma} = \boldsymbol{\sigma}^{\infty 1} + \boldsymbol{\sigma}^*, \quad \boldsymbol{\varepsilon} = \boldsymbol{\varepsilon}^{\infty 1} + \boldsymbol{\varepsilon}^* \tag{18}$$

where $\boldsymbol{\sigma}^{\infty 1}$ and $\boldsymbol{\varepsilon}^{\infty 1}$ are the stress and strain fields, respectively, induced by the dislocations (the infinite solution) with the entire domain $V$ having the same material properties of domain $V_1$ *(homogenous solution)*. Applying Hooke's law for each of the sub-domains, and using (18), one obtains the elastic constitutive equations for each of the materials in each of the sub-domains as:

$$\begin{aligned}\boldsymbol{\sigma}^* &= [\boldsymbol{C}_1^e]\boldsymbol{\varepsilon}^*, & in \ \ V_1 \\ \boldsymbol{\sigma}^* &= [\boldsymbol{C}_2^e]\boldsymbol{\varepsilon}^* + \boldsymbol{\sigma}^{\infty 21}; \ \ \boldsymbol{\sigma}^{\infty 21} = [\boldsymbol{C}_2^e - \boldsymbol{C}_1^e]\boldsymbol{\varepsilon}^{\infty 1}, & in \ \ V_2\end{aligned} \tag{19}$$

where $C_1^e$ and $C_2^e$ are the elastic stiffness tensors in $V_1$ and $V_2$, respectively. The boundary conditions are:

$$t = t^a - t_1^\infty \text{ on } S; \quad u = u^a \text{ on part of } S \tag{20}$$

where $t^a$ is the externally applied traction and $t_1^\infty$ is the traction induced on all of $S$ by the dislocations in $V_1$ in the infinite-homogenous domain problem. The "*eigenstress*" $\sigma^{\infty 21}$ is due to the difference in material properties. The method described above can be extended to the case of heterogeneous materials with $N$ sub-domains (Zbib and Diaz de la Rubia 2002)

The above system of equations (13-15) can be combined with the boundary corrections given by (16-20). The resulting set of field equations can be solved numerically using the finite element method as described by Zbib and Diaz de la Rubia (2002). The end result is a model, which they call a multiscale discrete dislocation dynamics plasticity (*MDDP*), coupling continuum elasto-visoplaticity with discrete dislocation dynamics. The *MDDP* consists of two main modules, the DD module and the continuum finite element module. The DD module computes the dynamics of the dislocations, the plastic strain field they produce, and the corresponding internal stresses field. These field values are passed to the continuum finite element module, in which the stress-displacement-temperature field is computed based on the boundary value problem at hand. The resulting stress field, in turn, is passed to the DD module and the cycle is repeated.

## 9.4 Typical Fields of Applications and Examples

Over the past decade, the discrete dislocation dynamics has been utilized by a number of researchers to investigation many complicated small-scale crystal plasticity phenomena that occur under a wide range of loading and boundary conditions (see for example a recent review by Bulatov et al. (2001)), and covering a wide spectrum of strain rates. Some of the major phenomena that have been addressed include:

- The role of dislocation mechanisms in strain hardening (Devincre and Kubin 1997; Zbib, Rhee et al. 2000).
- Dislocation pattern formation during monotonic and cyclic loading.
- Dislocation-defect interaction problems, including dislocation-void interaction (Ghoniem and Sun 1999), dislocation-SFT/void-clusters interaction in irradiated materials and the role of dislocation mechanisms on the formation of localized shear bands (Diaz de la Rubia, Zbib et al. 2000; Khraishi, Zbib et al. 2002).
- Effect of particle size on hardening in metal-matrix composites (Khraishi and Zbib 2002).
- Crack tip plasticity and dislocation-crack interaction (Van der Giessen and Needleman 2002).
- The role of various dislocation patterns such geometrically necessary boundaries (GNB's) in hardening phenomena (Khan, Zbib et al. 2001).
- Plastic zone and hardening in Nano-indentation tests (Fivel, Roberston et al. 1998).
- The role of dislocation mechanisms in increased strength in nano-layered structures (Zbib and Diaz de la Rubia 2002; Schwartz 2003).
- High Strain Rate Phenomena and shock wave interaction with dislocations.

In what follows, and in order to illustrate the utility of this approach in investigating a wide range of small-scale plasticity phenomena, representative results for a set of case studies are presented. This includes, dislocation behavior during monotonic loading, the evolution of deformation and dislocation structure during loading of a cracked specimen, and dislocation interaction with shockwaves during impact loading conditions[1]. The results given below are for both copper and molybdenum single crystals whose materials properties are as follows. For copper, mass density=8,900 kg/m$^3$, $G$ = 54.6 GPa, $\nu$ = 0.324, b = 0.256 nm, $M = 10^4$ 1/Pa.s. For molybdenum, mass density=10,200 kg/m$^3$, $G$ = 123 GPa, $\nu$=0.305, b = 0. 2725 nm, $M_{mixed}$= $10^3$/Pa.s, $M_{screw}$= 0.1/Pa.s.

---

[1] The dynamic evolution of the dislocation structures presented in this paper can be best visualized by viewing the video clips available on the website http://www.cmm.wsu.edu/

## 9.4.1 Evolution of Dislocation Structure during Monotonic Loading

A classical problem in crystal plasticity is the nature of strain hardening behavior and the underlying dislocation microstructure evolution during both loading and stress relaxation. Both patterning and the concomitant strain hardening are thought to result from attractive non-planar dislocation interactions that lead to formation of sessile junctions and intersections. The first simulation result shown is that of deformation of a single crystal with periodic boundary conditions under monotonic loading with low strain rate. For this case, one may constructs a simulation cell with either reflected boundary conditions (Hirth, Rhee et al. 1996; Zbib, Rhee et al. 1996) or periodic boundary conditions (Bulatov, Rhee et al. 2000), maintaining dislocation continuity and flux across the cell boundaries. In either case, the long-range stress field is computed using the super-dislocation method described in Section 9.2.6. Figure 3a shows a cube unit cell whose size is 10 μm with initial screw dislocations (with jogs and kinks) that are distributed randomly in the crystal. The crystal considered is molybdenum with randomly distributed initial dislocation density on the <111>{011} systems. The load is applied in the $[\bar{2}\,9\,20]$ direction for reasons described in (Lassila, LeBlanc et al. 2002). A constant strain rate of 1/s is imposed in that direction. At this strain rate the DD simulation is performed with the time step varied between $10^{-8}$s to $10^{-6}$s. The dislocation velocity and the shortest distance between two dislocations control the time step. The stress field is assumed to be uniform throughout the cell, and therefore, the finite element part of the analysis is suppressed. The number of steps in this simulation was over one million (one processor on Dec Alpha workstation) to reach a strain of about 0.3 %. Typical results can be seen in Figure 3b showing the dislocation morphology at 0.3 % strain. From the data collected, one can extract various interesting information that could be useful in many ways. For example, by analyzing the spatial distribution of all dislocation segments, one can construct pair-distribution functions as shown in Figure 4, for projections in various crystallographic directions, from which one can extract a wavelength and indication of a dislocation pattern. Further analysis also reveals that not only the <111>{011} systems are active but also some of the <111>{112} systems also become activated, resulting mainly from multiple cross-slip.

### 9.4.2 Dislocation Crack Interaction: Heterogeneous Deformation

Figures 5 and 6 show results of a mode I crack in a single crystal copper. The crack is located at the right side of the bottom of the specimen as can be deduced from the figures. The crack length is one third of specimen width. The bottom surface of the specimen (the un-cracked portion) and the left and right sides of the specimen are assumed to be symmetric boundaries. The crystal orientation is depicted in the figure with the x, y and z-axes being in the (110), ($\bar{1}10$) and (001) directions respectively. Initial dislocation loops (Frank-Read sources) are distributed randomly in the crystal on two slip planes: the ($\bar{1}11$) and ($1\bar{1}1$) planes. The initial dislocation density is $10^{12}$ m/m$^3$. The upper surface of the specimen is displaced a constant distance such that the over all macroscopic strain is 1.67 % (stress relaxation condition). This orientation induces initially double slip deformation, but as the simulation proceeds, some of the dislocations segment cross-slip and multi-slip deformation prevails. The DD simulations is performed in parallel with the finite element analysis which corrects for boundary tractions, image stresses from the surfaces of the crack, and computes the heterogeneous stress field.

Figures 5a depicts the distribution of shear stresses in the cracked specimen and after the dislocations attained the distribution shown in Figure 6a with the dislocation density having increased by about two orders of magnitudes to $7 \times 10^{13}$ m/m$^3$ (see Figure 6b). Figure 5b shows the corresponding plastic strain distribution. The stress field here does not have the exact same characters of the elastic stress field of a dislocation fee cracked-specimen. This is due to two factors, one is the internal stress field of the dislocations, and two is the plastic strain field induced by the mobile dislocations, both of these fields cause stress relaxation and reduce the order of the stress singularity at the crack tip. This can be deduced from Figure 5b where the plastic strain around the crack tip is about 1.5%. Close examination of Figure 6a reveals the formation of pile-ups of dislocation near a boundary separating the region with high stresses and a region with very low stress (almost insignificant). The dislocation piles up consist of [101] and [011] type dislocations extended in the (110) direction. The density of these pile ups is high enough that could lead to the nucleation of a microcrack that eventually would link with the main crack. The coupled DD-FE analysis also predicts the evolution of the temperature field. In this simulation the rise in temperature at these strain levels was about 20 K in a region where the plastic strain is about 3.5 % around the crack tip (Figure 5b).

### 9.4.3 Dislocations Interaction with Shock Waves

In this final example, we give representative results pertaining to the deformation of a pure copper single crystal under extreme pressures and high strain rates(Shehadeh and Zbib 2002). The main objective of these simulations is to investigate the interaction of dislocations with stress waves under extreme strain rates up to $10^8$/s and pressures ranging from a few GPa to tens GPa, under which the elastic properties are pressure-dependent. The simulation consists of a domain shown in Figure 7a. The cross-section is a square whose size 2.5μm. The length f the cell in the z-direction is 25 μm. The cell is considered to be in an infinite domain with the four side surfaces having symmetric boundary conditions, i.e. the FE nodes can slide in the plane of their respective surface but not normal to it . This condition is assumed in order to keep the specimen under confinement to better represent shock wave experiments (Meyers 1994; Kalantar et al. 2001). The upper surface is displaced (compressively) with a constant velocity over a small period of time (e.g. 3 ns) to correspond to an average strain rate of $10^6$/s, the surface is then released and the simulation continues long enough for the elastic waves to pass through the entire domain and dissipate at the lower boundary. During this process the dislocation sources, which are situated in the simulation cell as can be deduced from Figure 7a, interact with the stress wave and an avalanche of dislocations takes place the instant the wave impacts the sources. As the wave passes through the sources, dislocations continue to emit from the sources but eventually the rate of production diminishes and a dislocation structure with high density emerges as can be deduced from Figure 7b. Figure 8a depicts the pressure profile -calculated by the finite element analysis -along the length of the specimen after the wave has traveled past the dislocations sources. The figure also shows the wave profile when there are no dislocations present in the crystal. It can be deduced form the figure that the dislocations cause reduction in the pressure peak resulting from plastic strain and internal stress relaxation. The effect of the length duration of the imposed pulse on the dislocation density is shown in Figure 8b. Finally, the FE analysis also provides information about the temperature field as can be seen in Figure 9a. The increase in temperature is due to dissipation of energy from dislocation drag. Figure 9b shows the history profile of the temperature in the region with the highest rate of energy dissipation.

## 9.5 Summary and Concluding Remarks

The discrete computational approach of dislocation dynamics is a *bridging* model that is based on the fundamental carriers of plastic deformation -- dislocations. The DD approach developed over the past decade has overcome many hurdles and established a useful framework for predictive simulations. In order to expand the range of engineering and scientific issues that can be addressed, the DD models should become more realistic and computationally efficient. A much-needed extension of the DD methodology is to anisotropic elasticity and a consistent treatment of local lattice rotations that become increasingly important with increasing strain. A completely new set of challenging problems arises when dealing with high strain rates phenomena. When it comes to putting dynamics into dislocation dynamics, relevant extensions should incorporate relativistic distortions of the stress field of the fast moving dislocations and various peculiar behaviors associated with the short-range interactions.

Despite the impressive recent developments in the DD methodology, it is important to avoid unrealistic expectations. DD models were intended to address crystal plasticity at the microscopic length scale. Stretching their computational limits is important but may not be the most constructive way to building a predictive framework for macroscale plasticity. This is because deformation processes at still larger (meso) scales are important, involving collective behavior of huge dislocation populations and making direct DD simulations computationally prohibitive. Further coarse-graining and homogenization are required so that the results of DD simulations at the microscale are used to inform a less detailed meso-scale model. Although no consistent framework of this kind currently exists, several recent developments in this area look promising. In addition to the reaction-diffusion approaches (Holt 1970; Walgraef and Aifantis 1985), variant forms of strain gradient plasticity theory have been advanced attempting to capture length scale effects associated with small-scale phenomena (Mindlin 1964; Dillon and Kratochvil 1970; Aifantis 1984; Zbib and Aifantis 1989; Fleck, Muller et al. 1994; Arsenlis and Parks 1999). Closely related are several recent attempts (Anthony and Azirhi 1995; Shizawa and Zbib 1999) that follow the original ideas of Kröner (1958) and describe the populations of crystal defects as continuous fields generating geometric distortions in the crystalline lattice. One way or another, computational prediction of crystal strength presents the over-arching goal for micro- and meso-scale material simulations.

**Acknowledgement:** The support of Lawrence Livermore National Laboratory to WSU is gratefully acknowledged. This work was performed under the auspices of the U.S. Department of Energy by Lawrence Livermore National Laboratory (contract W-7405-Eng-48).

Khraishi, T. and Zbib, H. M., 2002. Dislocation dynamics simulations of the interaction between a short rigid fiber and a glide dislocation pile-up. Comp. Mater. Sci. 24, 310-322.

Khraishi, T., Zbib, H. M. and Diaz de la Rubia, T., 2001. The treatment of traction-free boundary condition in three-dimensional dislocation dynamics using generalized image stress analysis. Materials Science and Engineering A309-310, 283-287.

Khraishi, T., Zbib, H. M., Diaz de la Rubia, T. and Victoria, M., 2002. Localized deformation and hardening in irradiated metals: Three-dimensional discrete dislocation dynamics simulations,. Metall. Mater. Trans. 33B, 285-296.

Kocks, U. F., Argon, A. S. and Ashby, M. F., 1975. Thermodynamics and kinetics of slip. Oxford, Pergamon Press.

Koppenaal, T. J. and Kuhlmann-Wilsdorf, D., 1964. The effect of prestressing on the strength of neutron - irradiated copper single crystals. Appl. Phys. Lett. 4, 59.

Kröner, E., 1958. Kontinuumstheorie der versetzungen und eigenspanungen, Srpinger-Verlag.

Kubin, L. P., 1993. Dislocation patterning during multiple slip of FCC Crystals. Phys. Stat. Sol (a), 135, 433-443.

Kubin, L. P. and Canova, G., 1992. The modelling of dislocation patterns. Scripta Metall. 27, 957-962.

Lassila, D. H., LeBlanc, M. M. and Kay, G. J., 2002. Uniaxial stress deformation experiment for validation of 3-D dislocation dynamics simulations. ASME J. Eng. Mat. Tech. 124, 290-296.

Le, K. C. and Stumpf, H., 1996. A model of elasticplastic bodies with continuously distributed dislocations. Int. J. Plasticity 12, 611-628.

Lepinoux, J. and Kubin, L. P., 1987. The dynamic organization of dislocation structures: A simulation. Scripta Metall. 21, 833-838.

Mason, W. and MacDonald, D., 1971. Damping of dislocations in Niobium by phonon viscosity. J. Appl. Phys. 42, 1836.

McKrell, T. J. and Galligan, J. M., 2000. Instantaneous dislocation velocity in iron at low temperature. Scripta Materialia 42, 79-82.

Meyers, M. A., 1994. Dynamic behavior of materials. John Wiley & Sons, INC.

Mindlin, R. D., 1964. Arch. Rat. Anal., 16, 51.

Mügge, O., 1883. Neues Jahrb. Min 13.

Orowan, E., 1934. Zur Kristallplastizitat II. Z. Phys 89, 614.

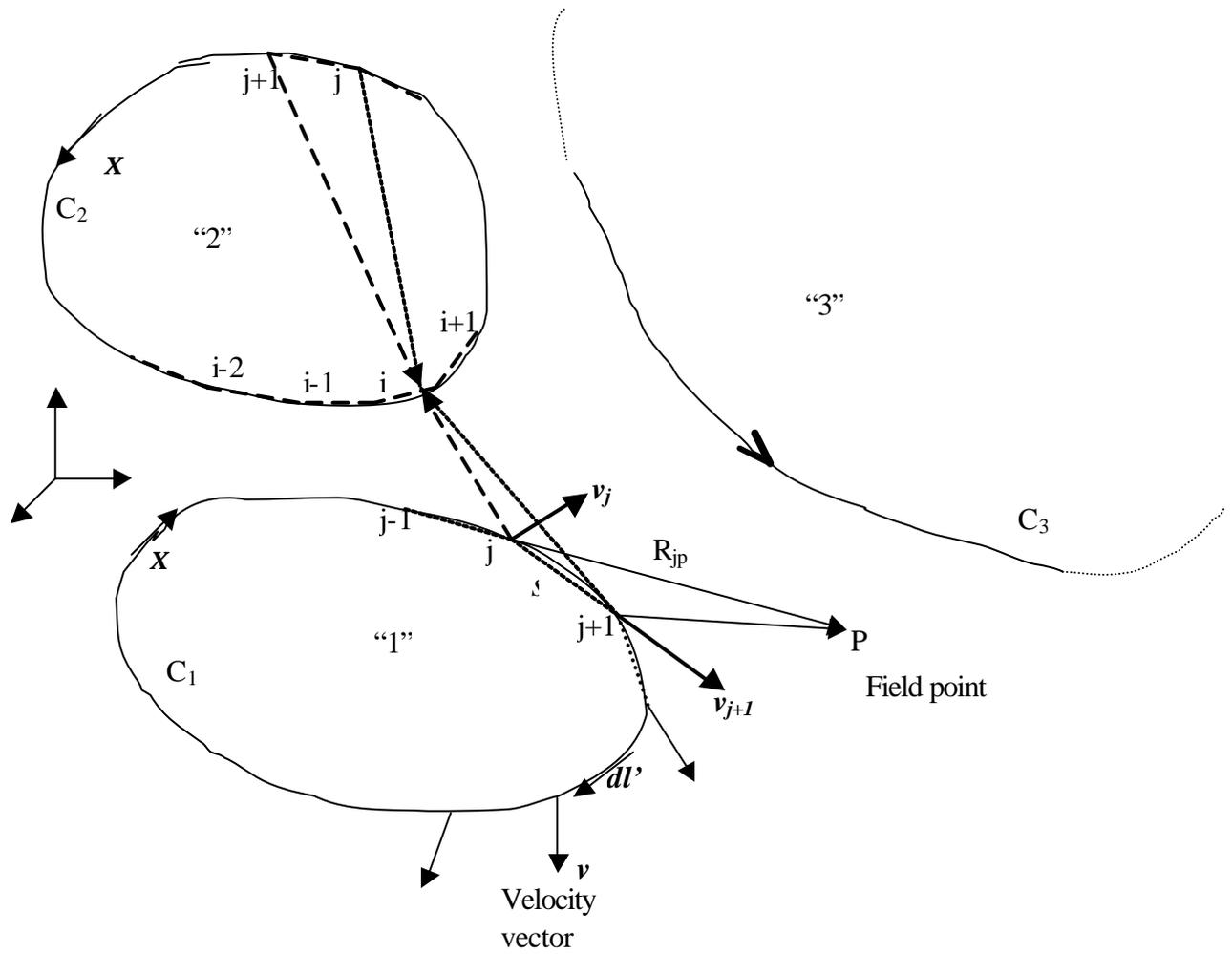

Figure 1. Discretization of dislocations loops and curves into nodes, segments and collocation points.

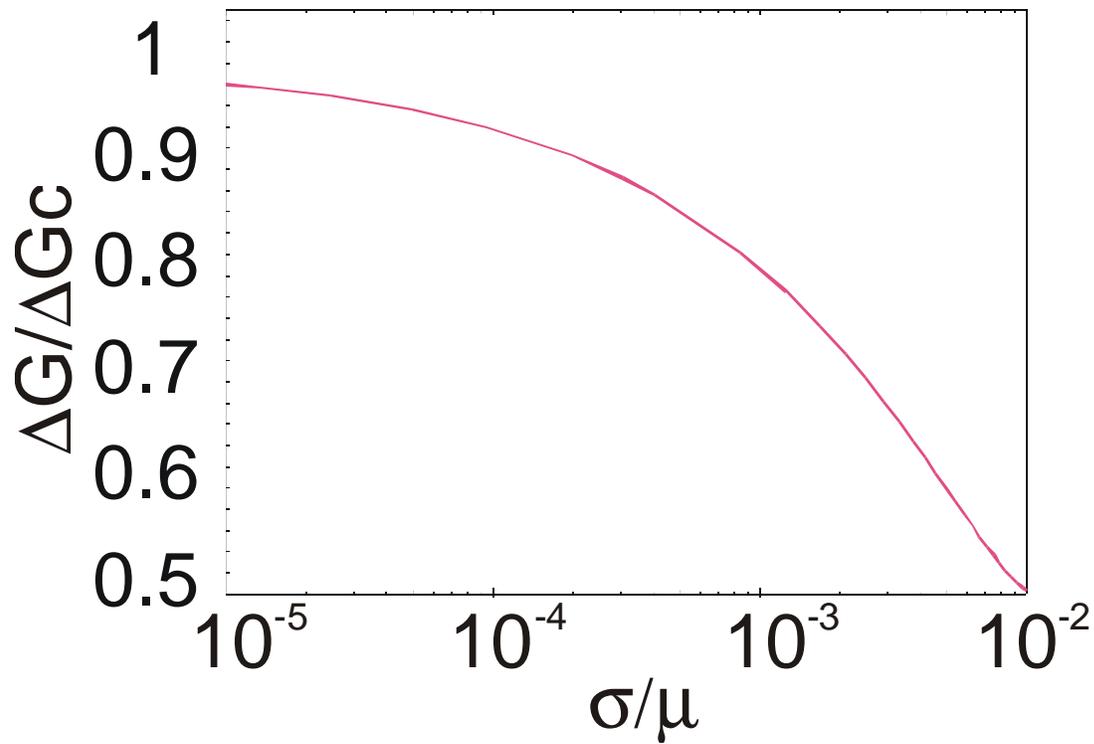

Figure 2. The normalized activation enthalpy for copper as a function of the normalized resolved shear stress on the glide plane. $G_c$ and $\mu$ denote the activation free energy and the shear modulus, respectively.

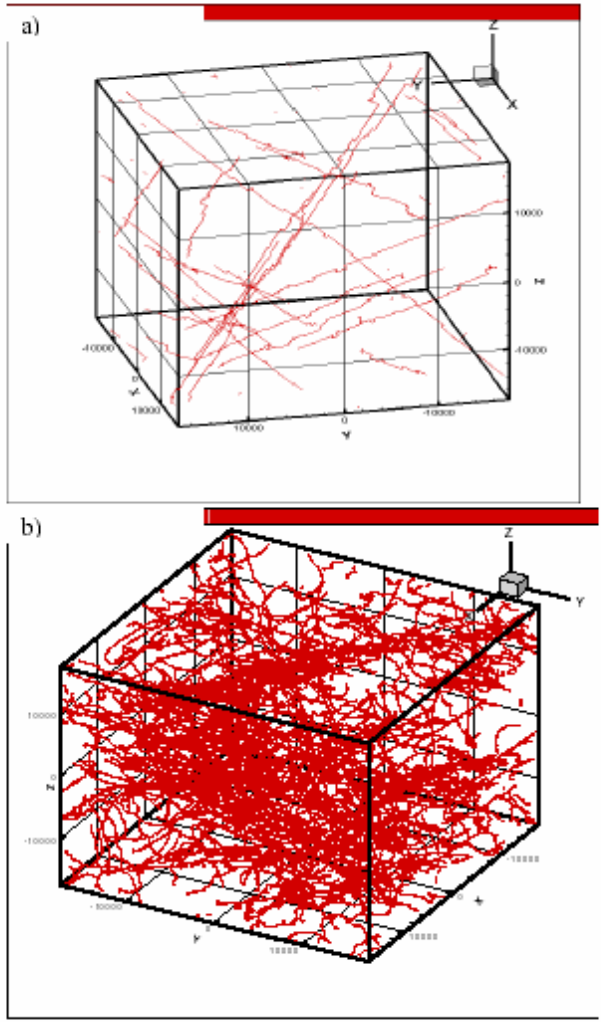

Figure 3. Deformation of molybdenum of single crystal under constant strain rate (1/s).
a) Initial dislocation structure, b) dislocation structure at 0.3% strain.

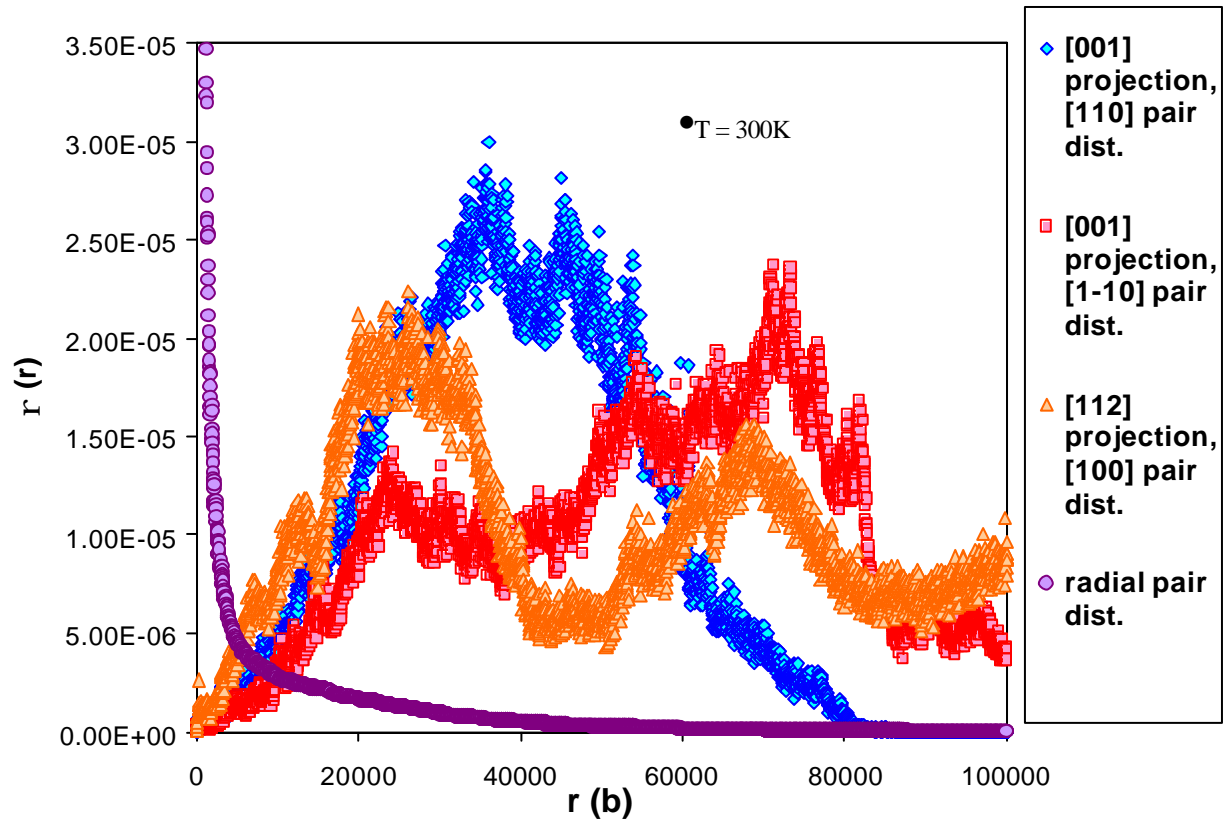

Figure 4. Pair distribution functions constructed for the dislocation structure shown Figure 1b

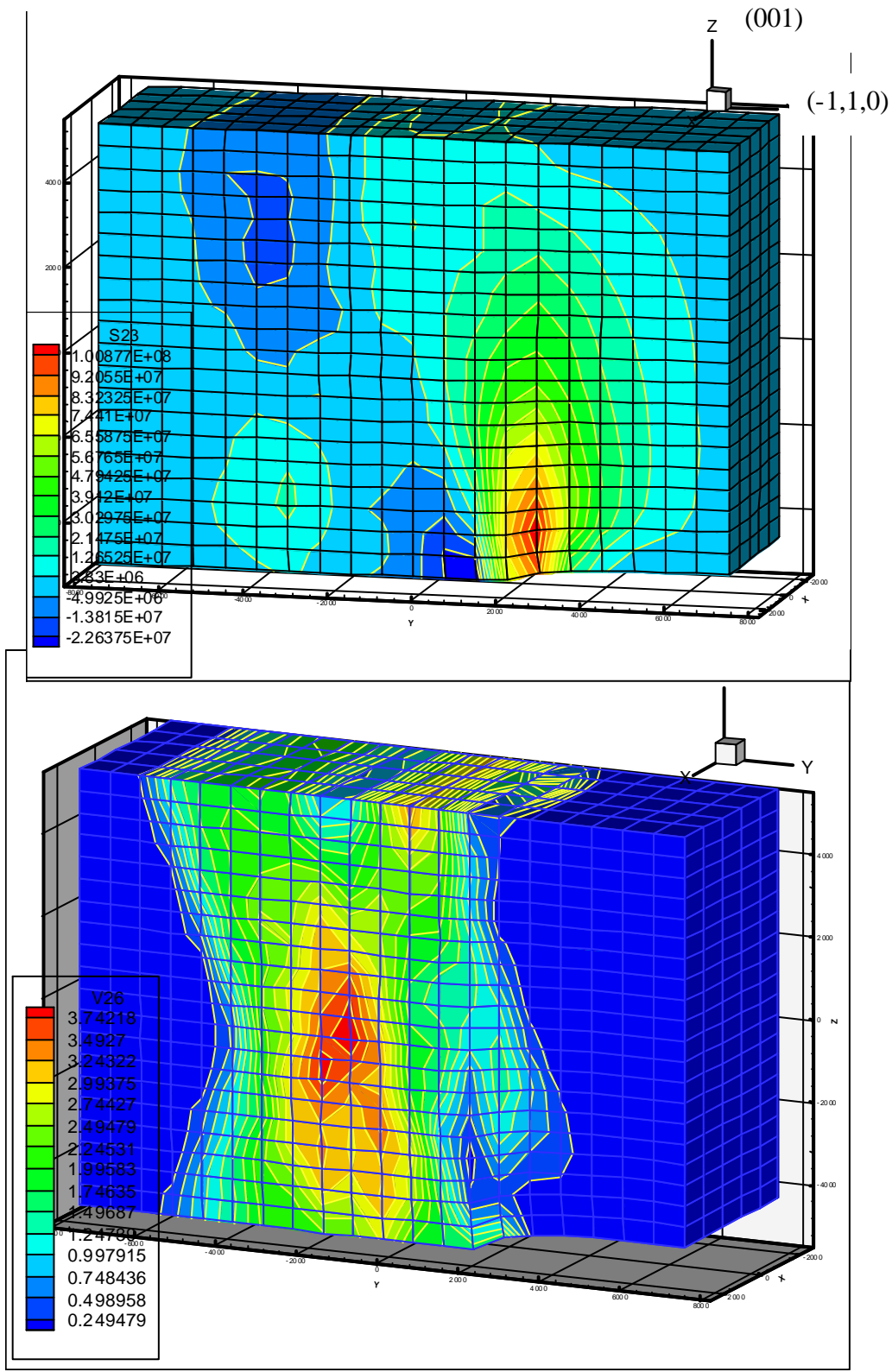

Figure 5. Coupled DD-FE analysis of a mode I crack. Top figure is a contour plot of the shear stress yz, and the bottom figure is a contour plot of the % effective plastic strain resulting from the dislocation motion.

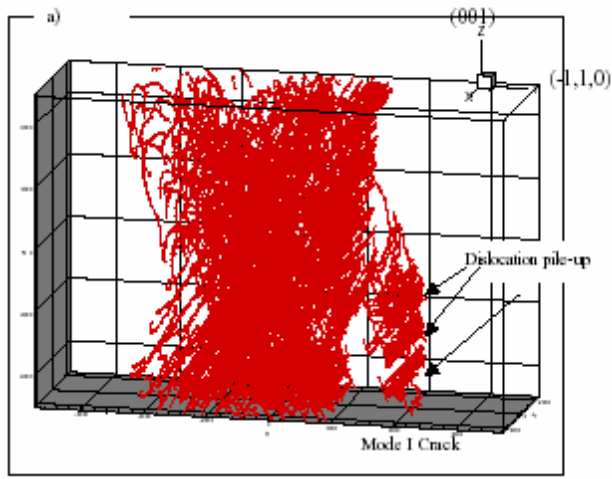

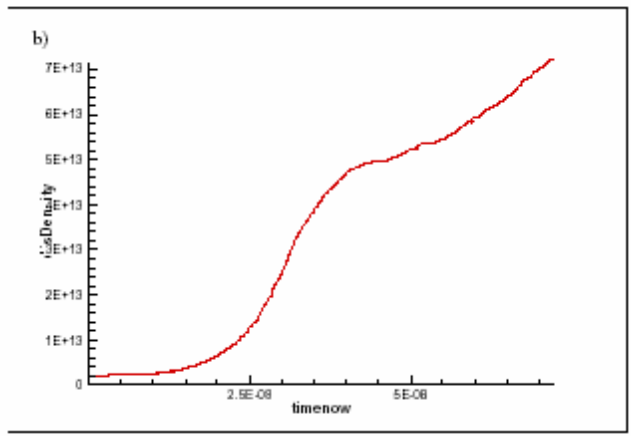

Figure 6. a) The dislocation structure around the crack tip. b) The evolution of the dislocation density.

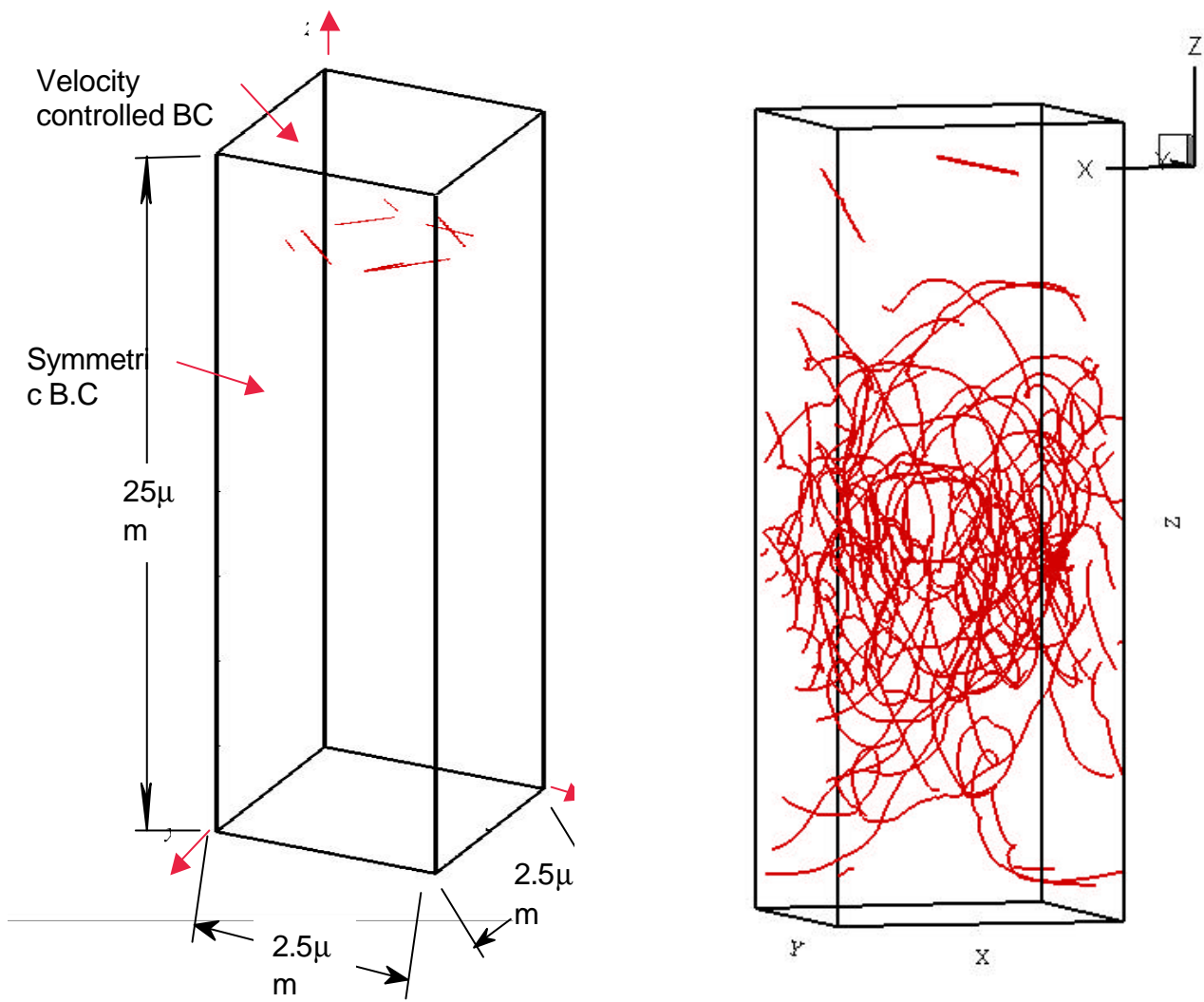

Figure 7. Dislocation-Shockwave interaction. a) Computational setup geometry with initial dislocation distribution (dislocation loops, Frank-Read sources). b) The dislocation morphology in crystal shocked at strain rate $1 \times 10^6$ for 3.6 nanoseconds

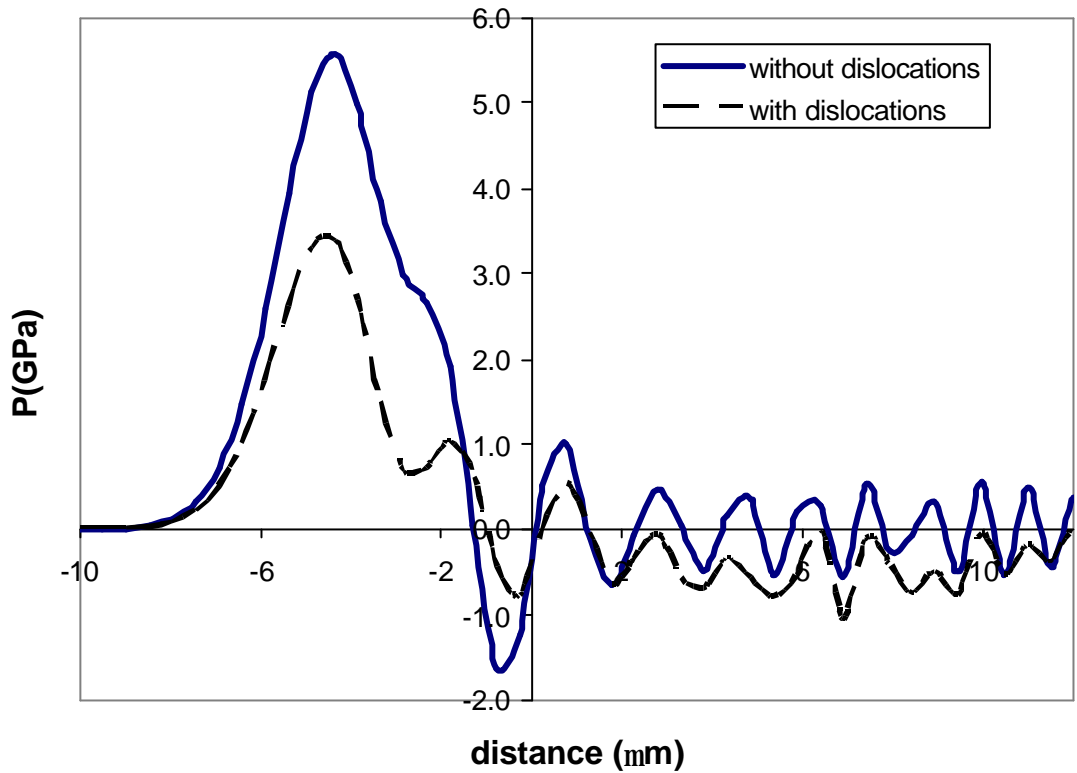

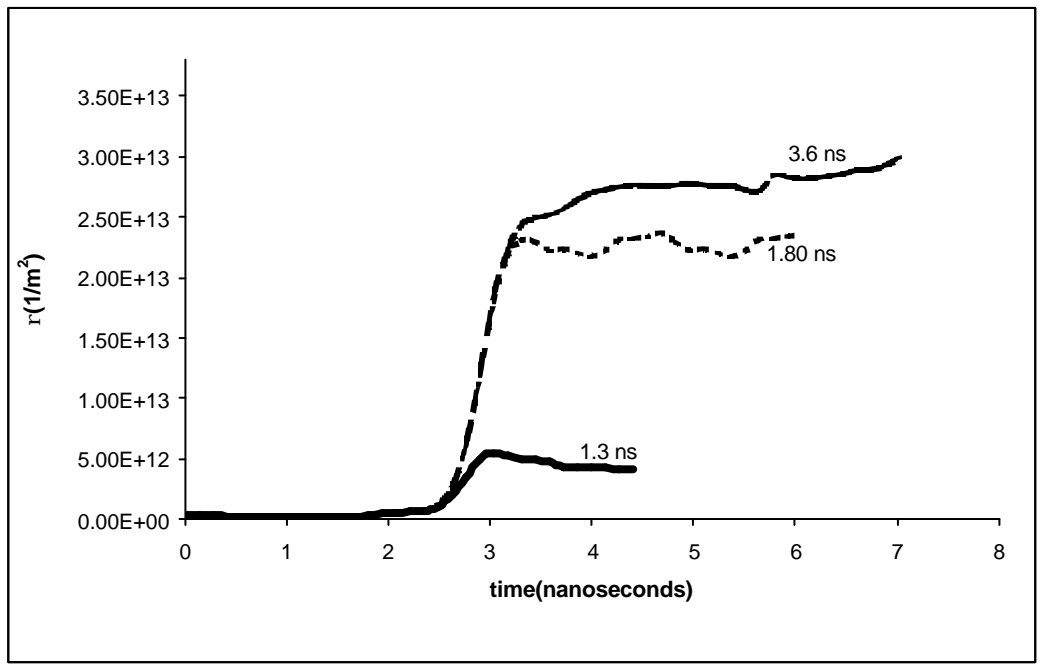

Figure 8. Dislocation-shockwave interaction. A) Effect of dislocations on pressure profile. B) Effect of pulse period on the dislocation density.

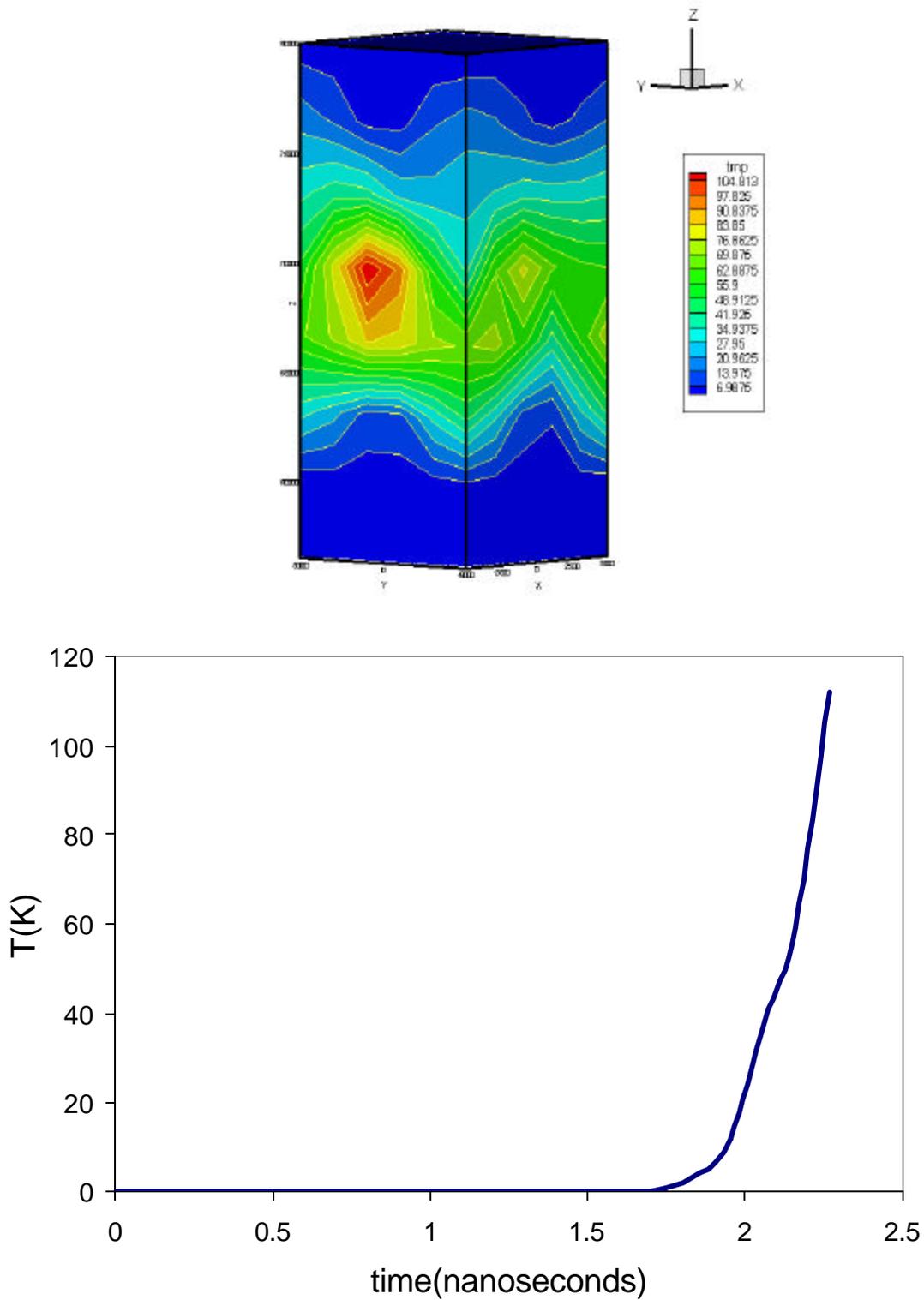

Fig. 9. a) Contour plot of temperature distribution in the computational cell, the local temperature increase reached value as high as 105 °C. b) The temperature history in the element of maximum temperature increase.